\documentclass[prb,preprint]{revtex4-1} 

\usepackage{amsmath}  
\usepackage{amsfonts} 
\usepackage{graphicx} 

\begin{document}

\title{Statistical Mechanical Interpretation of 2D Pauli Crystals}
\author{Aditya Mishra}
\email{aditya.mishra.phy@gmail.com}
\altaffiliation{ Department of Physics, Birla Institute of Technology, Mesra, Ranchi - 835215, India\\}
\author{Swarnava Mitra}
\email{swarnava.phy@gmail.com}
\altaffiliation{ Department of Physics, Birla Institute of Technology, Mesra, Ranchi - 835215, India\\}



\date{\today}

\begin{abstract}
A manifestation of the Pauli Exclusion Principle is observed when fermions are trapped in the ground state of a 2D harmonic oscillator trap at very low temperatures. This non-interaction of fermions results in the formation of Pauli crystals. This work introduces a statistical mechanical interpretation of the principles that led to the observation of Pauli crystals by calculating the energy, radius and other parameters of the non-interacting arrangement of ultra-cold fermions. This model approaches the problem from two different directions, namely fermion degeneracy and harmonic oscillator treatment. Unifying the two different approaches gives a more comprehensive and robust description of the various parameters of Pauli crystals.
\end{abstract}

\maketitle 

\section{Introduction} 

Fermions, i.e., particles with half-integer spin, are known to obey the Pauli Exclusion Principle (PEP) which is a consequence of the Fermi-Dirac statistics. PEP simply states that no two fermions with the same spin can ever exist in the same quantum state simultaneously. In their paper, Gajda et al. \cite{gajda16} had theorised crystal structures as a result of this \textit{non-interaction} between fermions due to the PEP and the resulting formation of structures due to their geometric arrangement. Although they are called Pauli crystals, the name itself is a misnomer. Pauli crystals have no translational symmetry and lack long range order- properties that are present in any crystal. Pauli crystals were first observed in May, 2020 by a group led by M. Holten from the University of Heidelberg, Germany. These crystals form an interesting topic for study and this paper presents an alternative point of view from the existing literature. The novelty of this paper lies in the statistical mechanical description of radius, energy and other parameters of the crystals. The results presented in subsequent sections match closely with the experimental values obtained by Holten et al. in their paper.~\cite{holten} 

Previous attempts have been made to come up with a model to explain the behaviour and properties of Pauli crystals.\cite{anal} This paper is an attempt to describe a more accurate theoretical interpretation of the higher order correlations arising due to PEP. Unlike the previous attempts, this paper has stressed on fermion degeneracy and used it as a stepping stone to derive the parameters. Pauli crystals have always been approached by the harmonic oscillator treatment. This paper takes a different approach and takes the fermion degeneracy route. Finally, unifying the two different approaches forms a robust and succinct description of Pauli crystals.

\section{Origin and imaging of Pauli crystals}
This section is based on the work done in Refs. \cite{holten,nandini,gajda17,chiu,prev,gajda20,gajda21}.
PEP states that no two identical fermions can ever occupy the same quantum state simultaneously. PEP applies to only fermions, bosons do not obey this; they have the freedom to crowd. A more appropriate mathematical and quantum mechanical approach can be presented with the use of the Slater determinant. 

Fermions are described by antisymmetric wave functions. An antisymmetric wave function can be mathematically described as follows:
\begin{equation}
\Psi(\mathbf{x}_1, \mathbf{x}_2) = -\Psi(\mathbf{x}_2, \mathbf{x}_1)
\end{equation}

To approximate the wave function of a two-particle case with coordinates $\mathbf{x}_1$ and $\mathbf{x}_2$, we have the Slater determinant. It can be expressed as
\begin{equation}
\begin{aligned}
 \Psi(\mathbf{x}_1, \mathbf{x}_2) &= \frac{1}{\sqrt{2}} \{\chi_1(\mathbf{x}_1) \chi_2(\mathbf{x}_2) - \chi_1(\mathbf{x}_2) \chi_2(\mathbf{x}_1)\} \\
  &= \frac{1}{\sqrt{2}} \begin{vmatrix} \chi_1(\mathbf{x}_1) & \chi_2(\mathbf{x}_1) \\ \chi_1(\mathbf{x}_2) & \chi_2(\mathbf{x}_2) \end{vmatrix},
\end{aligned}
\end{equation}
Where the coefficient is the Normalization factor. This wave function is now antisymmetric and no longer distinguishes between fermions. Moreover, it also goes to zero if any two spin orbitals of two fermions are the same. This is equivalent to satisfying the Pauli exclusion principle.

The expression can be generalised to any number of fermions by writing it as a determinant. For an N-electron system, the Slater determinant is defined as
\begin{equation}
\begin{aligned}
 \Psi(\mathbf{x}_1, \mathbf{x}_2, \ldots, \mathbf{x}_N) &=
  \frac{1}{\sqrt{N!}}
  \begin{vmatrix} \chi_1(\mathbf{x}_1) & \chi_2(\mathbf{x}_1) & \cdots & \chi_N(\mathbf{x}_1) \\
                      \chi_1(\mathbf{x}_2) & \chi_2(\mathbf{x}_2) & \cdots & \chi_N(\mathbf{x}_2) \\
                      \vdots & \vdots & \ddots & \vdots \\
                      \chi_1(\mathbf{x}_N) & \chi_2(\mathbf{x}_N) & \cdots & \chi_N(\mathbf{x}_N)
  \end{vmatrix} 
\end{aligned}
\end{equation}
The normalization constant is implied by noting the number N, and only the one-particle wavefunction is written.

As a consequence of the PEP, a few noninteracting fermions trapped by untracold lasers self-organize into a periodic structure called a Pauli crystal. Pauli crystals represent the most probable arrangement of fermions when they have to fill ground state energy levels in an N-fermion system after being cooled to low temperatures. These geometric structures are not symmetric as one would expect 'crystals' to be (at least in the 2D case) and lack the regularity of 2D polyhedral structures. Pauli Crystals are unique in this way.

The N-fermion system should ideally be a very weakly interacting system. In case of an interacting system, the crystals would not be visible. For an N-particle system, as mentioned before, the wave function can be determined by the Slater determinant. Here too the same can be done and will be given by
\begin{equation}
\Psi(r_1, ..., r_N ) = \sqrt{\frac{1}{N!}} \det{[\psi_i(r_k)]}
\end{equation}

The single particle wave functions of an oscillator bound in a 2D harmonic trap is given by
\begin{equation}
\Psi_{p,q}(x,y)=\frac{e^{-(\frac{x^2+y^2}{2}) }H_p(x)H_q(y)}{\sqrt{2^{p+q}p!q!\pi}}
\end{equation}
where $H_p$ is the $p^{th}$ order Hermite polynomial and $x,y$ are in harmonic oscillator units. The single-particle energy corresponding to these states is $E_{pq}=(p+q+1)\hbar\omega=(n+1)\hbar\omega$ where $n=p+q$ is the total excitation number and each energy level is $(n+1)$-fold degenerate. This wave function in turn gives the probability density function which can be maximised using the Monte-Carlo algorithm. This gives us the theoretical arrangement of the fermions in the crystal.

The experimental detection and imaging of a Pauli crystal has two important parts. The first one is N non-interacting ultracold fermions trapped in a 2D harmonic potential and the second one is single shot imaging of the resulting geometric structure in terms of either position or momentum. Each single shot of the system of fermions determines the existence of a particular correlation. This process is repeated to recover all such correlations formed by the particles. Due to the rotational symmetry being broken with each individual experiment, the images obtained are found to be rotated by some angle with respect to an arbitrary axis. All N images have slightly different angular orientation so they will have to be rotated and then averaged together about the centre of mass of the system to reveal the correlations that we term Pauli crystals.  
Holten et al. \cite{holten} have taken the particle momentum into consideration to determine the geometries arising out of the Pauli crystal and their experiment presents a practical showcase of the PEP.

\section{Total energy from fermion degeneracy}\label{sec1}
In this section, we derive the total energy of Pauli crystals from the fermion degeneracy approach. 
\subsection{Energy of the crystal}
Consider a fermion is captured in a 2-D box of sides L. This box can also be thought of as an area of zero potential surrounded by walls of infinitely high potential. The fermion cannot
penetrate infinitely high potential barriers.

Solving the Schrödinger equation for a 2-D well we get,
\begin{equation}
 \psi_{n_x n_y}=\frac{2}{L}\sin{\left(\frac{n_x\pi}{L}x\right)}\sin{\left(\frac{n_y\pi}{L}y\right)}  
\end{equation}

The corresponding energy eigenvalues are:
\begin{equation}
E_{n_x n_y}=\frac{\pi^2\hbar^2}{2mL^2}(n_x^2+n_y^2)
\end{equation}

Fermions with the highest allowed energy in this system will be at the Fermi Energy $E_f$
\begin{equation}
E_f=\frac{\pi^2\hbar^2}{2mL^2}R^2
\end{equation}
Where $R^2=n_x^2+n_y^2$ is the radius in R-space.

Therefore, number of energy eigen value points in R-space is given by
\begin{equation}
N=\pi R^2\times2\times\left(\frac{1}{2}\times\frac{1}{2}\right)=\frac{\pi}{2}R^2
\end{equation}
The factor of 2 is multiplied for the two spins and the halves are introduced to remove redundant $\pm n_i$.\\
Hence the value of $R^2$ becomes,
\begin{equation}
R^2=\frac{2}{\pi}N
\end{equation}
And, the value of $E_f$ becomes,
\begin{equation}
E_f=\frac{\pi^2\hbar^2}{2mL^2}\left(\frac{2N}{\pi}\right)=\frac{\pi\hbar^2}{m}\left(\frac{N}{L^2}\right)
\end{equation}
The number density of the fermions is given by 
\begin{equation}
\sigma=\frac{N}{L^2}=\frac{N}{V}
\end{equation}
Where $V=L^2$ has been chosen to represent the area; analogous to the volume for 3-D systems.\\
Substituting the value of $\sigma$ in eqn (6) we get
\begin{equation}
E_f=\frac{\pi\hbar^2}{m}\sigma
\end{equation}
The density of states for a 2-D system is given by
\begin{equation}
g(E)=\frac{m}{\pi\hbar^2}
\end{equation}
This can be easily derived using a treatment similar to what we have done above. However, it can be concluded logically that the density of states for a 2-D system is independent of the energy. Since, as soon as some energy is provided to the particles, a significant number of states are available to them.

\vspace{5mm}
The total energy can be computed by integrating the combined energy of all the states below the state that has the Fermi energy
\begin{equation}
U=\int_{0}^{E_f} E.g(E)  \,dE
=\frac{mL^2}{\pi\hbar^2}\frac{E_f^2}{2}=\frac{mL^2}{2\pi\hbar^2}\left(\frac{\pi\hbar^2}{m}\right)^2 \sigma^2 = \frac{\pi\hbar^2}{2m}\sigma^2L^2
\end{equation}
The equation can be rearranged in the form
\begin{equation}\label{e1}
U=\frac{\pi\hbar^2}{2m}\left(\frac{N^2}{V}\right)
\end{equation}
Eqn (\ref{e1}) gives the total energy of a Pauli crystal.

\subsection{Energy in harmonic oscillator units}
We use natural harmonic oscillator units. 
The linear unit of the ground state wavefunction in a cartesian direction is given by $a_0 =\sqrt{\hbar/m\omega}$.\cite{pathria} 

Energy is in units of $E_0 =\hbar \omega$. 

Converting Eqn (\ref{e1}) into natural harmonic oscillator units would somewhat simplify it and make it more comprehensive. Also, the data in Holten et al.~\cite{holten} is given in units of harmonic oscillator. This makes it easier for comparison with experimental values.

Let the radius of the crystal in harmonic oscillator units be $r$. 

Therefore, the actual radius of the crystal $R=ra_0$.

Thus, the actual area of the crystal becomes
\begin{equation}
V=\pi R^2=r^2\frac{\pi\hbar}{m\omega}
\end{equation}

According to Eqn (\ref{e1}), the total energy is given by
\begin{equation}
U=\frac{\pi\hbar^2}{2m}\left(\frac{N^2}{V}\right)=\frac{\pi\hbar^2}{2m}\left(\frac{N^2}{r^2\frac{\pi\hbar}{m\omega}}\right)=\frac{N^2}{2r^2} \hbar\omega
\end{equation}
Therefore, in harmonic oscillator units the total energy of a Pauli crystal is given by 
\begin{equation}\label{e2}
U=\frac{N^2}{2r^2}
\end{equation}
Where $N$ is the total number of particles in the crystal and $r$ is the radius of the crystal in harmonic oscillator units.

\subsection{Comparison with experimental data}

The total number of fermions $N=6$ and the radius is $r\approx 1.14$. The exact figure was not reported in Ref. \cite{holten}. This value has been taken from Figure 2c of Ref. \cite{holten}.

According to Eqn (\ref{e2}), the total energy of the crystal (in units of $E_0$) turns out to be
\begin{equation}
U=13.85
\end{equation}
The theoretical value in Ref. \cite{holten}(in units of $E_0$) is 
\begin{equation}
U=14
\end{equation}
This is very close to the value from our calculations with an error of only $1.1\%$. The error stems solely from the value of the radius.
\\
And, the experimentally determined value is 
\begin{equation}
U=13.1
\end{equation}
The experimental value deviates from our calculated value by about $7.1\%$.

It must be noted that the paper itself quoted an experimental error of $6\%$. 

One can be quite apprehensive of the utility of this treatment of a Pauli crystal. This only gives an approximate value that is also dependent on the measured radius of the crystal system. Moreover, treating the fermions in the crystal as harmonic oscillators would eliminate the above two drawbacks. However, if we have some patience we shall observe this will help us to predict the radius as well as the energy independent of each other.

\section{Total energy from harmonic oscillator treatment}
In this section, we derive the total energy of Pauli crystals by treating the fermions as harmonic oscillators. To ease out this process, we define a quantity called the order of the crystal. 
\subsection{Order of crystal}
Due to the Pauli exclusion principle, fermions are filled into the energy levels in an ordered fashion. They progress in series of natural numbers as 1,2,3,...

Therefore, total number of fermions in a Pauli crystal is of form 
\begin{equation}
N=1+2+3+...+k=\frac{k(k+1)}{2}    
\end{equation}
Where, $k$ is a natural number i.e., $k \in \mathbb{N}$ .

Here $k$ can be deemed as the order of the Pauli crystal that gives us the total number of fermions in the crystal.
\begin{figure}[h!]
\centering
\includegraphics[width=0.9\textwidth]{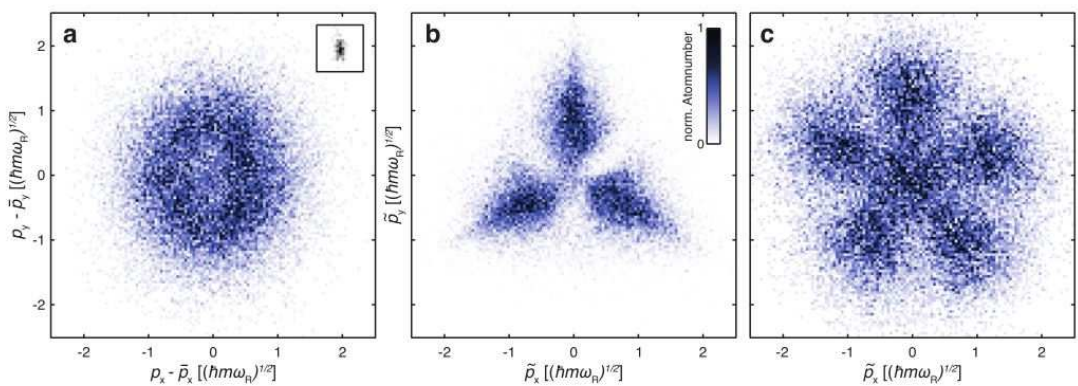}
\caption{Crystals of (a) order 1  (b) order 2  (c) order 3. Image from Ref. \cite{holten}}
\end{figure}

For a crystal of order 1 the total number of fermions is 1 i.e., only one at the center. For a crystal of order 2 the total number of fermions is 3, all residing in one single orbit. For a crystal of order 3, total number of fermions is 6 with one at the centre and the other five in an orbit around the centre.

\subsection{Energy of the $k$-th order crystal}
The energy of the single particle states is given by $E=(n+1)\hbar\omega$, where $n$ is total
excitation number of the degenerate energy levels. The $n$-th energy level is $(n + 1)$-fold degenerate.

For the total energy of the $k$-th order or $N=\frac{k(k+1)}{2}$ ground state, we need to sum over the energies of all $N$ fermions. Keeping the mind the degeneracy, the $n$-th energy level can be occupied by $(n + 1)$ fermions.

The result is as follows :

1 fermion in the n=0 level : $E=1\hbar\omega$

2 fermions in the n=1 level: $E=(2\times2)\hbar\omega=4\hbar\omega$

3 fermions in the n=2 level: $E=(3\times3)\hbar\omega=9\hbar\omega$

\vspace{5mm}
Total energy for $N=1+2+...+n=\frac{n(n+1)}{2}$ fermions is $[1^2+2^2+3^2+...+n^2]\hbar\omega$.

\vspace{5mm}
Clearly, $\frac{n(n+1)}{2}=\frac{k(k+1)}{2}$ implies $n=k$.

\vspace{5mm}
Therefore, the total energy of the $k$-th order
\begin{equation}\label{e3}
U=(1^2+2^2+3^2+...+k^2)\hbar\omega=\frac{k(k+1)(2k+1)}{6}\hbar\omega
\end{equation}
Rewriting Eq (\ref{e3}) in terms of $N$ and in harmonic oscillator units, we get 
\begin{equation}\label{e4}
U=\frac{N}{3}\sqrt{8N+1}
\end{equation}
Despite containing a square root and division by 3, U will always be an integer since it is the sum of the squares of natural numbers.

\subsection{Comparison with data}
The following energies are in harmonic oscillator units.

From Eq (\ref{e4}), the total energy of a 2nd order crystal with $N=3$ is $U=5$.

The total energy of a 3rd order crystal with $N=6$ is $U=14$.

The total energy of a 4th order crystal with $N=10$ is $U=30$.

\section{Radius of a Pauli crystal}
From Eq (\ref{e2}) and (\ref{e4}),
\begin{equation}
\frac{N^2}{2r^2}=\frac{N}{3}\sqrt{8N+1}
\end{equation}
Therefore, the radius in harmonic oscillator units
\begin{equation}\label{e5}
r=\left(\frac{3N}{2\sqrt{8N+1}}\right)^\frac{1}{2}
\end{equation}
In Eq (\ref{e5}), the radius depends on one variable only. We can calculated the radius of crystal by knowing only the number of fermions it contains.

Rewriting Eq (\ref{e5}) in terms of order $k$
\begin{equation}
r=\sqrt{\frac{3k(k+1)}{4(2k+1)}}
\end{equation}

For $N=3$, the radius $r=0.949$. 

For $N=6$, the radius $r=1.134$. This is quite close to the  data from Ref. \cite{holten}.

In non-natural units, Eq (\ref{e5}) becomes
\begin{equation}
R=r\sqrt{\frac{\hbar}{m\omega}}=\left(\frac{3\hbar}{2m\omega}.\frac{N}{\sqrt{8N+1}}\right)^\frac{1}{2}
\end{equation}

\section{Density of a Pauli crystal}
In Sec. (\ref{sec1}), the number density of the fermions in a Pauli crystal has been denoted as $\sigma=N/V$. 

Therefore, in natural units 
\begin{equation}\label{e6}
\sigma={\frac{N}{\pi r^2}}=\frac{2}{3\pi}\sqrt{8N+1}
\end{equation}

In non-natural units,
\begin{equation}
\sigma=\frac{2m\omega}{3\pi\hbar}\sqrt{8N+1}
\end{equation}

\section{Number of fermions in a shell}\label{sec10}

Here shell refers to the circuital arrangement of the fermions in the crystal; not to be confused with energy levels or orbitals.

In a Pauli crystal, the fermions arrange themselves in a way  maximizing the N-body probability distribution. This is usually calculated using the Monte Carlo algorithm. However, we can take note of the fact there is a particular pattern to the arrangement of these fermions as shown in Fig. 1 of Ref. \cite{gajda17}.
\begin{figure}[h!]
\centering
\includegraphics[width=0.9\textwidth]{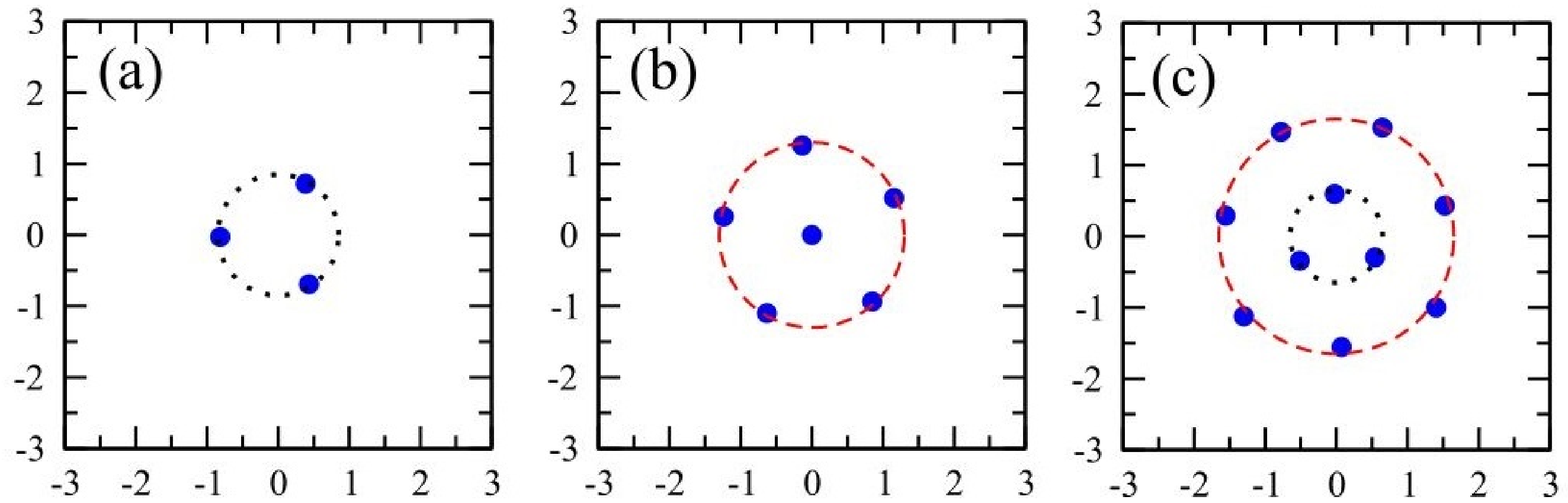}
\caption{Pauli crystals of (\textbf{a}) order 2 (\textbf{b}) order 3 (\textbf{c}) order 4. The even ordered crystals progress as $1,5,9,..$,    and the odd ordered crystals progress as $3,7,11,..$. Image from Ref. \cite{gajda17}}
\end{figure}
The number of fermions in the $i$-th orbit of a crystal or $k$-th order, can be expressed as
\begin{equation}
f_i=4i+2(k\ mod\ 2)-3
\end{equation}
Where $mod$ is the modulo function that gives the \textit{least positive residue} (i.e., remainder of Eucledian division). $(a\ mod\ b)$ gives remainder when $a$ is divided by $b$.

For even $k$, $(k\ mod\ 2)=0$. Therefore, $f_i=4i-3$ or the number of fermions in orbits progresses as $1,5,9,..$.

For odd $k$, $(k\ mod\ 2)=1$. Therefore, $f_i=4i-1$ or the number of fermions in orbits progresses as $3,7,11,..$.

\section{Predictions for $N=15$ fermions}
Let us try applying the above model to a system of $N=15$ fermions. $N=15$ is a crystal of order 5. The following values are in harmonic oscillator units. 

From Eq (\ref{e4}), the total energy of a 5th order crystal is $U=55$.

From Eq (\ref{e5}), the radius of the crystal is $r=1.43$.

From Eq (\ref{e6}), the number density of the crystal is $\sigma=2.33$

From Sec. (\ref{sec10}), the number of fermions in the first, second and third shells are $1,5\ \&\ 9$ respectively.

\begin{figure}[h!]
\centering
\includegraphics[width=0.6\textwidth]{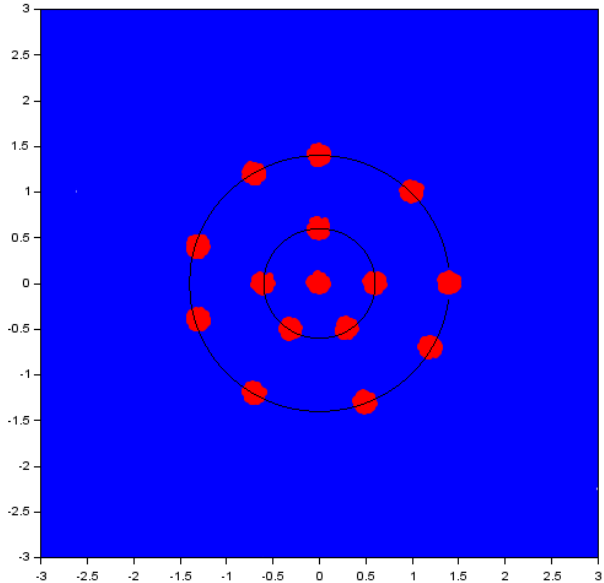}
\caption{Monte-Carlo simulation for $N=15$ fermions}
\label{7png}
\end{figure}

The Monte-Carlo simulation for the same has been done in \textbf{Fig.\ref{7png}}. From the simulation, it is crystal clear that the structure matches our predictions.

\section{Conclusion}
We made predictions for a system of $N=15$ fermions based on the model developed in this paper. The results match closely with the results obtained from the Monte-Carlo simulations in \textbf{Fig.\ref{7png}}.

Having said that, we would like to reiterate that the objective of this work is to develop a simpler model for studying the  qualitative and quantitative properties of the Pauli crystals, keeping the statistical mechanical treatment in mind. This model approaches the problem from two different directions, namely fermion degeneracy and harmonic oscillator treatment. As a result, this yields much more precise and accurate results compared to previous attempts.

\medskip
\medskip

\begin{acknowledgments}
The authors would like to acknowledge Dr Sourabh Lahiri of  Birla Institute of Technology, Mesra on two grounds. Firstly, for inculcating in his students an interest in statistical mechanics that made this work possible. Secondly, his endorsement has made it possible for this work to be uploaded into arXiv. The authors would also like to express their gratitude towards Dr Madhu Priya of Birla Institute of Technology, Mesra for taking an exceptional foundational course in statistical mechanics and assigning a project that resulted in this work. AM would also like to thank his mother for being a constant source inspiration for him.
\end{acknowledgments}

\end{document}